\documentstyle[11pt,paspconf,epsf]{article}
 \def\mso{\, M_\odot}                                                           

 \def\lso{\, L_\odot}                                                           

 \def\kms{\, {\rm km}\, {\rm s}^{-1}}

 \def\ra{\rightarrow}

 \def\h1{\hangindent=1.0truecm \hangafter=0}

 \def\simle{\mathrel{\hbox{\rlap{\hbox{\lower4pt\hbox{$\sim$}}}\hbox{$<$}}}}    
 \def\simgr{\mathrel{\hbox{\rlap{\hbox{\lower4pt\hbox{$\sim$}}}\hbox{$>$}}}}    
 \def\msoy{\, \mso~{\rm yr}^{-1}}                                               

\begin{document}

\title{The Evolution of Surface Parameters of Rotating Massive Stars}
\author{Norbert Langer}
\affil{Institut f\"ur Theoretische Physik und Astrophysik,
   Universit\"at Potsdam, D--14415 Potsdam, Germany}
\author{Alexander Heger}
\affil{Max-Planck-Institut f\"ur Astrophysik, D--85740 Garching, Germany}

\begin{abstract}
We summarize the present status 
of the predictions of massive star models for the evolution of
their surface properties. After discussing luminosity, temperature and
chemical composition, we focus on the question whether massive stars
may arrive at critical rotation during their evolution, either on the
main sequence or in later stages. We find both cases to be possible
and briefly discuss observable consequences.
\end{abstract}


\section{Introduction}
Massive main sequence stars are rapid rotators, with equatorial rotation
velocities in the range of $100 ... 400\kms$ (Fukuda 1982, Penny 1996,
Howarth et al. 1997). It is known since a long time that rotation can
affect the stellar interior in several ways. {\it Rapid rotation} can reduce
the effective gravity in the star, and it produces large scale flows
(Eddington 1925). During the evolution, {\it differential rotation} 
occurs in all stars, with the possibility of the occurrence of various local
hydrodynamic instabilities (cf. Endal \& Sofia 1978, Zahn 1983)
and corresponding mixing of chemical elements and angular momentum.
Of relevance for massive stars are the shear instability (cf. Maeder 1997),
the baroclinic instability (Zahn 1983, Spruit \& Knobloch 1984),
and the Solberg-H{\o}iland and Goldreich-Schubert-Fricke instabilities
(cf. Korycansky 1991). 

Time dependent evolutionary models for massive stars including rotation
have been constructed in the past in one dimension, using various degrees
of approximation (e.g. Endal \& Sofia 1978, Maeder 1987, Langer 1991,
Talon et al. 1997, Langer 1998). 
Today, it is beyond reasonable doubts that the evolution of massive stars
is influenced by rotation due to the physical mechanisms mentioned
above (cf. also Fliegner et al. 1996). While the principle effects of
rotation in the interior of massive stars during their evolution all the
way to iron core collapse are described elsewhere (Langer et al. 1997a,
Heger et al. 1998a), we concentrate here on observable surface parameters,
i.e. (latitudal averages of) luminosity, effective temperature and
surface abundances (Sect.~2), and equatorial rotation velocity.
In particular, we discuss the question whether massive stars have the
potential to evolve their surface to critical rotation, either during
core hydrogen burning (Sect.~3) or beyond (Sect.~4). 

\section{Evolution of luminosity, surface temperature, and abundances}
Fig.~1 displays the main effects of rotation on the initial
position and evolution of massive stars in the HR diagram at the
example of 10$\mso$ tracks for various degrees of rotation
(cf. Fliegner et al. 1996). First, the centrifugal force reduces the
effective gravity in the stellar interior, i.e. the star appears to
be less massive. Its luminosity and surface temperature are reduced 
(von Zeipel 1924, Kippenhahn 1977). The order of magnitude of this
effect can be seen comparing the ZAMS positions of the 10$\mso$ tracks.

However, during the further evolution of core hydrogen burning the effect
of chemical mixing becomes dominant. Shear instability and Eddington-Sweet
currents transport chemical elements synthesized in the stellar interior 
outwards, while the baroclinic instability 
smoothes out chemical gradients on equipotential surfaces. Due to the transport
of helium into the envelope the average mean molecular weight of the star is
increased compared to the non-rotating case, leading to much higher
luminosities (Kippenhahn and Weigert 1990). 

The effect on the surface temperature depends on the amount of mixing,
i.e. on the degree of rotation. In the extreme case of chemically
homogeneous evolution, the stars would evolve to the left of the ZAMS
directly towards the helium main sequence (cf. Maeder 1987). However,
more typical may be the case of moderate rotation in Fig.~1,
which brings the star to cooler surface temperatures than the non-rotating
models (cf. also Langer 1991). I.e., the main sequence band may be
considerably widened due to rotationally induced mixing, which  may
make the requirement of ``convective core overshooting'' 
(Stothers \& Chin 1992, Schroeder et al. 1997) obsolete. 

In any case, Fig.~1 shows that even on the main sequence the stellar
evolutionary track in the HR diagram depends on the initial rotation rate.  
I.e., rotation does not only have quantitative effects but qualitatively
alters fundamental stellar characteristics as isochrones, the initial
mass function, and mass-luminosity relations (Langer et al. 1997b).

A similar statement holds for the surface composition of massive stars:
it is altered stronger for larger initial masses but also for larger 
initial rotation rates. In principle, all chemical species which are 
affected by proton captures at core hydrogen burning temperatures can
show variations at the surface of rotating stars. However, as shown by
Fliegner et al. (1996), the variations of different species do occur at 
different times. For example, boron is depleted very early during the
main sequence evolution, while nitrogen and helium enrichments are
achieved only much later. Fliegner et al. use~B and~N observations
in B~stars to show that the abundance pattern in massive early
type stars (cf. Venn et al. 1996, and references therein)
is in fact produced by rotational mixing and not by close binary interaction.

\begin{figure}
\epsfxsize=0.8\hsize
\centerline{\epsffile{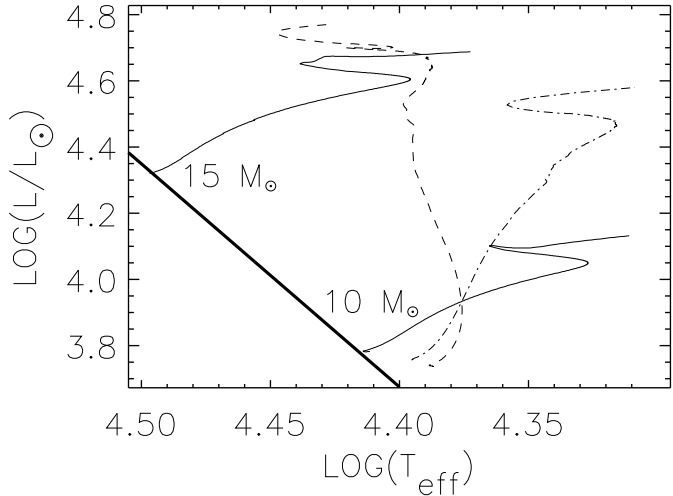}}
\caption{Evolutionary tracks in the HR diagram for
non-rotating 10 and 15$\mso$ stars (solid line), and a moderately
(dashed-dotted) and rapidly rotating (dashed) 10$\mso$ star, during the
main sequence evolution. The thick solid line marks the ZAMS position for
non rotating models (cf. Fliegner et al. 1996).}
\end{figure}

The time sequence of element abundance alteration is boron depletion,
nitrogen enhancement together with carbon depletion,
oxygen depletion, helium enhancement, and possibly sodium enhancement. 
The radionuclide $^{26}$Al may also be transported to the surface
of rotating massive main sequence stars. For the effect of rotation
on isotopic chemical yields of massive stars see Langer et al. (1997a).

\section{Evolution of the rotational velocity during core hydrogen burning}
The evolution of the surface rotation rate of stars depends on three
processes: the expansion or contraction of the star during its evolution,
angular momentum redistribution due to the physical processes mentioned
in Sect.~1, and the loss of angular momentum at the stellar surface.

\begin{figure}
\epsfxsize=0.4\hsize
\centerline{\epsffile{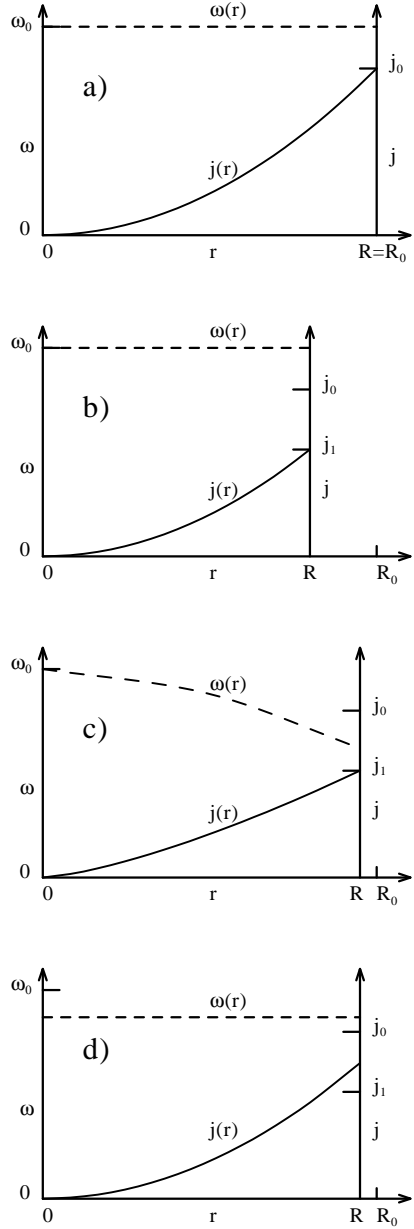}}
\caption{Schematic diagram of coupled mass and angular momentum
loss in a rigidly rotating star, broken up into three discrete
processes: mass loss without readjustment
($a) \ra b)$);
reexpansion of the star (($b) \ra c)$);
and reestablishment of rigid rotation (($c) \ra d)$).
$\omega$ is the angular velocity, $j=\omega r^2$ the specific
angular momentum, and $R$ the stellar radius.
Subscript~$0$ refers to the initial state.}
\end{figure}

During the main sequence evolution, the radius of massive stars increases
by a factor of 2...3. In case the specific angular momentum would remain
constant at the surface, the rotational velocity would decrease by 
that factor. However, according to Zahn (1994), rigid rotation is a good
approximation for the angular momentum distribution of massive main
sequence stars (however, see Maeder, this volume). In that case, the
transport of angular momentum out of the convective core --- which increases
its density by a factor of 2...3 during core hydrogen burning ---
supplies angular momentum for the surface layers such that, as net effect, 
their rotational velocity remains roughly constant (e.g., Packet et al. 
1980).

However, massive main sequence stars can lose angular momentum through a
stellar wind, even in the absence of magnetic fields. The mechanism of
this angular momentum loss is sketched in Fig.~2 for the case of rigid
rotation; it works in the same way for differentially rotating stars
provided that the time scale for angular momentum transport from the
core to the surface is shorter than the mass loss time scale.
Note that the effect of chemical evolution of the star, which leads to
an increase of the stellar radius with time, is neglected in Fig.~2.

Since stars of 10...20$\mso$ lose only small amounts of their total mass
during core hydrogen burning, they could be spun down only through
magnetic winds. However, main sequence mass loss may be substantial
for higher initial masses. Examining the evolution of 60$\mso$ stars,
Langer (1998) finds that massive main sequence 
stars may reach the $\Omega$-limit,
i.e. the state of critical rotation, with the critical rotational
velocity defined as to include the effect of radiation pressure
(cf. Langer 1997). The considered stars 
may reach critical rotation not by spinning up
but by a reduction of their critical rotational velocity as they evolve
closer to the Eddington limit. 

It is shown by Langer (1998) that massive main sequence stars may reach
the $\Omega$-limit without catastrophic consequences. Only the mass loss
rate is increased such that the corresponding angular momentum loss
rate (cf. Fig.~2) ensures that the $\Omega$-limit is never exceeded.
For a 60$\mso$ star, mass loss rates of the order of
$10^{-5}\msoy$ are achieved at the $\Omega$-limit, resulting in a
considerable spin-down. As the mass loss will not occur in a spherically
symmetric wind but rather in a disk, and since it is unclear whether
the stellar radiation can push all lost material to infinity
(cf. Owocki \& Gayley 1997), stars at the $\Omega$-limit might appear
peculiar, perhaps like B[e]~stars (Zickgraf et al. 1996).

\section{Evolution of the rotational velocity beyond core hydrogen exhaustion}
During the post main sequence evolution, strong chemical composition
and entropy gradients at the location of the hydrogen burning shell source
inhibit efficient mixing of angular momentum 
from the core into the hydrogen-rich envelope. Therefore, the angular
momentum evolution of the latter can be --- as first approximation ---
considered as independent of the core evolution (Heger et al. 1998ab).

Very massive stars may reach the $\Omega$-limit again immediately after
core hydrogen exhaustion. While the opacity peak which brought them
close to the Eddington limit on the main sequence is due to metal
opacities, the peak due to helium ionisation becomes relevant for
$T_{\rm eff} \simle 25\, 000\,$K. Since the stellar evolution proceeds
more than hundred times faster in this phase, correspondingly higher
mass loss rates have to be expected, with the result of a
more eruptive phenomenon, perhaps resembling Luminous Blue Variables
(Garc\'{\i}a-Segura et al. 1997).

\begin{figure}
\epsfxsize=0.7\hsize
\centerline{\epsffile{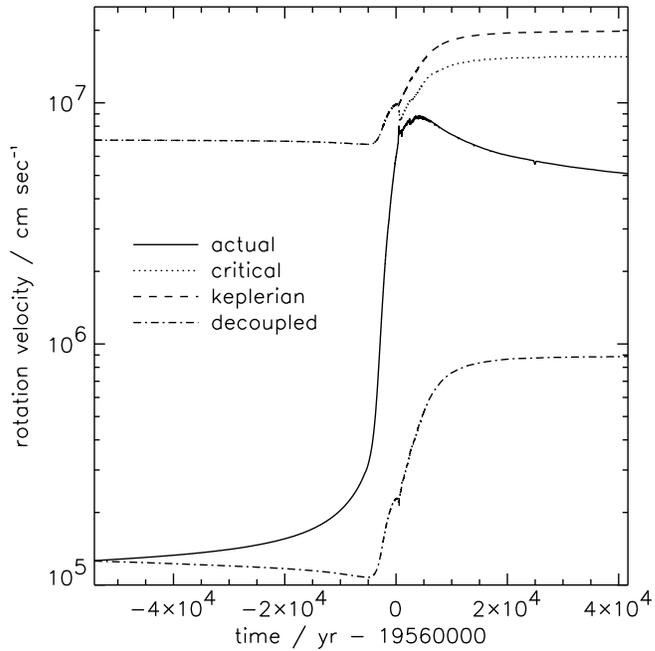}}
\caption{Equatorial rotation velocity as a function of time
  (solid line) of a rotating 12$\mso$ model during the transition
  from the red supergiant branch to the blue supergiants stage
  during core helium burining ($t=0$ is arbitrary).
  It is compared to the Keplerian (dashed line) and the
  critical rotation rate (dotted line); the latter two
  are different by the factor $1-\Gamma$, $\Gamma$ being the Eddington
  factor. During the red supergiant phase
  it is $\Gamma \ll 1$ and the two lines
  coincide, while during the blue supergiant phase $\Gamma$ rises to $0.4$.
  The dash-dotted line shows the evolution
  of the surface rotation
  rate if there were no angular momentum transport
  in the convective envelope (cf. Heger \& Langer 1998).}
\end{figure}

Heger \& Langer (1998) found that also stars with masses below
$\sim 20\mso$ may arrive at the $\Omega$-limit during their post-main
sequence evolution. Stars in that mass range may undergo so called
blue loops during core helium burning, i.e. excursions from the
red supergiant branch into the B~star regime. Since blue loops
are connected to a decrease of the stellar envelope by roughly a factor
of~10, a corresponding spin-up of the star may be expected. However,
at the same time the structure of the hydrogen-rich envelope changes
from convective to radiative. Heger \& Langer showed that the 
assumption of rapid angular momentum transport in convective regions
results in the fact that most of the angular momentum is retained in
the convective outer part of the envelope as the mass of this 
convective part decreases with time while retaining its spatial extent. 
Consequently, it spins up much more
than if angular momentum were conserved locally.

Fig.~3 gives an impression of the order of magnitude of the effect.
For the example of a 12$\mso$ star, the rotational velocity is increased
by a factor of~100 during the blue loop, where only a factor of~10 would
be expected from the mere contraction. This is enough to bring the star
to critical rotation, and only a mass loss enhancement and a correspondingly
high angular momentum loss rate prevents the star from exceeding the
$\Omega$-limit.

The consequence of this spin-up of contracting red supergiant envelopes
is an episode of highly aspherical mass loss. Its consideration may
be relevant for the interpretation of B[e] stars of rather low
luminosity ($\sim 10^4\lso$, cf. Gummersbach et al. 1995), for the
circumstellar material around blue supergiants (cf. Brandner et al. 1997)
in general, and Supernova 1987A in particular (Woosley et al. 1997),
and for post-AGB stars and the formation of bipolar planetary nebulae 
(Garc\'{\i}a-Segura et al. 1998).

\acknowledgments
This work was supported in part through the Deutsche Forschungsgemeinschaft
through grant La~587/15-1.

\end{document}